# Increasing biological complexity is positively correlated with the relative genome-wide expansion of non-protein-coding DNA sequences

Ryan J. Taft[1, 2]* and John S. Mattick[3]

[1]Rowe Program in Genetics, Department of Biological Chemistry, University of California, Davis, School of Medicine; [2]Current address: California Pacific Medical Center Research Institute, San Francisco, California; [3]ARC Special Research Centre for Functional and Applied Genomics, Institute for Molecular Bioscience, University of Queensland, Brisbane, Qld 4072, Australia. *Corresponding author: ryantaft@cooper.cpmc.org.

**Background:** Prior to the current genomic era it was suggested that the number of protein-coding genes that an organism made use of was a valid measure of its complexity. It is now clear, however, that major incongruities exist and that there is only a weak relationship between biological complexity and the number of protein coding genes. For example, using the protein-coding gene number as a basis for evaluating biological complexity would make urochordates and insects less complex than nematodes, and humans less complex than rice.

**Results:** We analyzed the ratio of noncoding to total genomic DNA (ncDNA/tgDNA) for 85 sequenced species and found that this ratio correlates well with increasing biological complexity. The ncDNA/tgDNA ratio is generally contained within the bandwidth of 0.05 – 0.24 for prokaryotes, but rises to 0.26 – 0.52 in unicellular eukaryotes, and to 0.62 – 0.985 for developmentally complex multicellular organisms. Significantly, prokaryotic species display a non-uniform species distribution approaching the mean of 0.1177 ncDNA/tgDNA ($p=1.58 \times 10^{-13}$), and a nonlinear ncDNA/tgDNA relationship to genome size ($r=0.15$). Importantly, the ncDNA/tgDNA ratio corrects for ploidy, and is not substantially affected by variable loads of repetitive sequences.

**Conclusions:** We suggest that the observed noncoding DNA increases and compositional patterns are primarily a function of increased information content. It is therefore possible that introns, intergenic sequences, repeat elements, and genomic DNA previously regarded as genetically inert may be far more important to the evolution and functional repertoire of complex organisms than has been previously appreciated.

## Background

The completion of the human genome project has introduced a quandary that has yet to be satisfactorily resolved. Until recently, the estimated number of protein-coding genes in the human genome was predicted to range from as low as 40,000 to as high as 120,000. However, it is now apparent that humans have no more than 30,000 protein-coding genes [1, 2], similar to other vertebrates such as the mouse [3] and pufferfish [4].

As the level of complexity of an organism increased, it was assumed that the number of genes would also proportionately increase. Specifically, it was stated that the number of distinct genes that an organism made use of was a valid measure of its complexity [5]. This must be true in the broad sense of the amount of encoded genetic information, but it is also dependent on the definition of a gene, which may be incomplete. Genes are usually considered to be synonymous with proteins, apart from those genes encoding infrastructural RNAs that are required for mRNA processing and translation (rRNAs, tRNAs, small nucleolar RNAs, and spliceosomal RNAs) and some that produce other non-protein-coding RNAs. Genomes are currently described in terms of their protein-coding gene capacity, on the expectation that proteins necessarily act out most, if not all, vital cellular functions. Non-protein-coding sequences are usually regarded as either cis-acting regulatory elements acting at the DNA or RNA level, or as evolutionary detritus. Although substantially more complex organisms have more protein-coding genes than simple ones, it is clear that the data are unable to validate the hypothesis that the numbers of protein-coding genes equate with biological complexity. By these criteria, insects and urochordates are less complex than nematodes, which have more genes, and mammals, despite their organizational complexity, are no more complex than plants or puffer fish. This observation has recently been named the gene number, or g-value, paradox by Hahn et al. [6]. While some of these inconsistencies and incongruities may be explained by alternative splicing, this may not be the entire explanation. It is becoming clear that complex multicellular organisms have a high degree of conservation of non-protein-coding DNA (ncDNA) elements and express large numbers of noncoding RNAs (ncRNAs) [7, 8].

Likewise, in the pre-genomic era multiple analyses indicated that there was no definable relationship between another measure of genetic information, the amount of DNA per cell, and biological complexity. For example, some protozoans, plants and amphibians contained more DNA per cell than mammals. This phenomena has been named the C-value enigma or paradox [6, 9]. However, these analyses could not take into account relative ploidy of these organisms, accurately assess the number of genes, or investigate general genomic architecture.

In order to examine biological complexity in light of the gene number and C-value paradoxes we have investigated the ratio of non-protein-coding DNA to total genomic DNA (ncDNA/tgDNA) in 85 sequenced genomes. The data suggest that general increases in biological complexity are positively correlated with increasing ncDNA/tgDNA ratios, albeit within a bandwidth influenced by variable amounts of repetitive sequences (of uncertain functional significance). Based on these results and a brief review of the current literature we suggest that intronic, intergenic and other genomic sequences previously regarded as "junk" [10], "gene deserts" [1], or "gene bare" [4] may be far more important to the functional repertoire and evolution of complex organisms than has been previously appreciated.

## Results and Discussion

### *Biological Complexity*

The definition of biological complexity is a matter of perspective and discussion, if not debate, and ranges from informational to phenotypic parameters. Attempts have been made to define biological complexity in a number of ways, including relationship to formal language theory, the parts considered, using thermodynamics, and most recently genetic information [6, 11-14]. While these definitions have been instrumental to furthering the discussion on biologic complexity they have not resulted in a scientific consensus. However, biological complexity has been widely accepted to be a function of the range of subcellular structures (prokaryotes versus eukaryotes), increasing numbers of cell types, organ structures, the functional repertoire of the organism, neural and immune function, and the intricate developmental processes necessary for the generation of these characteristics. The recent advent of the genomic era, however, has shifted discussions of complexity to genomic composition. Perhaps the most valuable genomic definition of biological complexity stems from an information theoretic approach. This definition suggests that an organism's complexity is a reflection of the physical complexity of its genome, i.e. the amount of information a sequence stores about its environment [14, 15].

For the purposes of this study we have considered biologic complexity a synthesis of the popularly accepted and information theoretic definitions. That is, increasing complexity must be a product of changes in both macroscropic characteristics indicating greater sophistication, and increases in information rich DNA sequences. Since these sequences are most likely not genes, it is possible that this information is stored in the noncoding portion of the genome. The ncDNA/tgDNA trend we report here refines previous definitions by introducing divisions between groups of organisms as would be expected using the popular definition of biological complexity (e.g. using subcellular structures), and by helping to explain the growing body of evidence indicating the information rich nature of non-protein-coding sequences.

*Analysis of sequenced genomes*

Currently, the genomes of human (*Homo sapiens*), mouse (*Mus musculus*), puffer fish (*Fugu rubripes*), a urochordate (*Ciona intestinalis*), fruit fly (*Drosophila melanogaster*), mosquito (*Anopheles gambiae*), round worm (*Caenorhabditis elegans*), two subspecies of rice (*Oryza sativa* L. ssp. *indica* and *Oryza sativa* L. ssp. *japonica* ), mustard plant (*Arabidopsis thaliana*), a fungus (*Neurospoa crassa*), two species of malarial parasite (*Plasmodium falciparum* and *Plasmodium yoelii yoelii*), the agent of sleeping sickness (*Tyrpanosoma brucei*), fission yeast (*Schizosaccharomyces pombe*), baker's yeast (*Saccharomyces cerevisiae*), and a multitude of prokaryotes have been completely sequenced, or almost completely sequenced, with many others en route. For the purposes of this study genomes from species spanning all three domains of life (59 bacteria, 8 archaea, and 18 eukaryotes - 7 simple eukaryotes, 1 fungus, 3 plants, 3 invertebrates, 1 urochordate, and 3 vertebrates), have been compared and contrasted against one another. The analysis was carried out by assessing the amount of known and predicted protein-coding sequences per haploid genome in relation to the measured total haploid genome size, derived from information in the relevant genome sequence publications and subsequent literature detailing the genomic composition of the organism of interest . The non-protein-coding sequences therefore include structural elements of chromosomes (centromeres, telomeres, origins of replication, matrix attachment regions etc.), intergenic sequences, introns in protein coding genes, cis-regulatory sequences operating at the DNA or RNA level (transcriptional promoters, enhancers, and 5'- and 3'UTR regulatory elements in mRNA), noncoding RNA genes (many of which also contain introns) [16], pseudogenes, repetitive sequences, and sequences which may act as spacers between functional elements.

*ncDNA/tgDNA ratios rise with organismal complexity*

The data shows that the ratio of noncoding DNA to total genomic DNA increases as biological complexity increases (Figure 1). In brief, the ratio of noncoding DNA to total genomic DNA (ncDNA/tgDNA) rises from 0.05 - 0.24 in prokaryotes to 0.26 – 0.52 in developmentally simple unicellular eukaryotes like yeast and Plasmodium, followed by the fungus *Neurospora crassa* with a value of 0.62, a range of 0.71 – 0.80 for plants, 0.74 – 0.93 for invertebrates, a urochordate with a value of 0.87, and finally to a range of 0.89 – 0.98 for vertebrates. Notably, both prokaryotes with the lowest values have been detailed as evolutionarily unique species. The bacteria, *T. maritime*, has been described as belonging to one of the deepest and most slowly evolving lineages in the eubacteria [17]. Likewise, the archaea, *N. equitans*, is suggested to be a genomically stable parasite that diverged anciently from the archaeal lineage [18]. Humans, on the other hand, hold the highest ncDNA/tgDNA value and may be reasonably considered to be the most complex organism in the biosphere, in terms of the combination of sophistication of body plan and neural capacity.

The data also identify two significant boundaries: the first between nucleate and enucleate species; and the second between unicellular and multicellular species. All prokaryotes (eubacteria and archaea) examined have ncDNA/tgDNA values less than 0.25, while all eukaryotic species have values greater than 0.25. This seems especially remarkable in light of the fact that there is no observed correlation between gene number or genome size and delineation between nucleate and enucleate organisms. The second boundary defines the upper limit of the unicellular / developmentally simple eukaryotic species examined, and separates these species from multicellular organisms. All unicellular eukaryotes examined have ncDNA/tgDNA ratios that fall in a discrete band from 0.25 to a current limit of 0.52, while all multicellular eukaryotes examined have ncDNA/tgDNA values greater than 0.62, although there are certainly extant species which fall between these figures. As has been observed with the previous boundary between enucleate and nucleate species, gene number and genome size are unable to group these organisms in a fashion consistent with their understood relationships to one another. Using the ncDNA/tgDNA ratio, however, we are able to observe genomic relationships that are much better correlated with known macroscopic, phylogenic, and morphologic similarities between species.

*ncDNA/tgDNA ratios in prokaryotes*

The species included in this analysis represent over half of all available sequenced prokaryotic genomes. The highest value is associated with Rickettsia *prowazekii*, excluding *Mycoplasma leprae*, which may be a special case of a genome in decay with many remnant protein-coding sequences [19]. It is nearly impossible to associate prokaryotes with differing levels of complexity, but we assume that differing levels of non-protein-coding sequences in these species reflect differing levels of regulatory sophistication imposed by the demands of their environment. Therefore, it is interesting to note that many species of the same genus have ncDNA/tgDNA values that place them adjacent to one another, or adjacent to other species identified as closely phylogenetically related (Table 1). Additionally, examining the species density of prokaryotes per ncDNA/tgDNA value reveals a distinct pattern. The mean ncDNA/tgDNA value for the 67 prokaryotic species examined is 0.1177, which is in agreement with earlier estimates that the majority of prokaryotic genomes contain 6-14% noncoding DNA [20]. However, the distribution of species shows an unexpected concentration approaching the mean (Figure 2). Analysis using a chi squared test, under the assumption of uniform distribution, suggests that the observed ncDNA/tgDNA data are highly unlikely by chance ($p = 1.58 \times 10^{-13}$). Additionally, there is an obvious nonlinear relationship between ncDNA/tgDNA ratio values and genome size ($r = 0.15$), which contradicts the dogmatic prediction that the relative amount of non-protein-coding sequences increase as genome size increases, as gene number appears to [18]. These data suggest the possibility that the noncoding regions of prokaryotic genomes may be evolutionarily and biologically constrained, presumably to encode cis-regulatory elements, notwithstanding the limited numbers of noncoding RNA genes recently discovered in bacteria [21]. If noncoding sequences are information rich it is possible that too few of these sequences may constitute insufficient information for a prokaryote to regulate necessary biologic processes. For example, the lowest nDNA/tgDNA value, 0.05, may be a necessary minimum for prokaryotes, and ncDNA/tgDNA values below 0.05 may be occupied by other organisms, such as viruses. Likewise, as the relative density of these sequences increase it may be difficult for prokaryote species to maintain and utilize noncoding regions without the sub-cellular structures present in eukaryotes. Interestingly, it has recently been shown that regulatory proteins increase as a quadratic function of genome size in prokaryotes, indicating that prokaryotic complexity may ultimately be limited by regulatory overhead [22].

*ncDNA/tgDNA ratios in eukaryotes and resolution of incongruities*

The calculated ncDNA/tgDNA values for the sequenced eukaryotic species increase in a fashion that correlates well with biological complexity. The noncoding DNA/tgDNA ratio analysis also helps to solidify a genomic compositional relationship between yeast species, and between other unicellular eukaryotic species. *S. pombe* has only 4,824 protein-coding genes [23], less than many bacteria which (e.g. *P. aeruginosa, M. loti and S. coelicor*) have a substantially higher number of genes (5,570, 6,752 and 7,825 respectively) [24-26]. Its ncDNA/tgDNA value, however, places the fission yeast in the same group as *S. cerevisiae* and other unicellular eukaryotic species. Tellingly, analysis of the *S. pombe* genome sequence indicated that "the transition from prokaryotes to eukaryotes required more new genes than did the transition from unicellular to multicellular organization" [23]. It was also, however, clearly accompanied by an expansion in noncoding DNA sequences, which cannot be rationalized on the basis that slower-growing eukaryotes can tolerate superfluous DNA as has been suggested in the past [27, 28]. Specifically, some yeasts, such as *S. cerevisiae* and *S. pombe*, can have generation times similar to or faster than many prokaryotes, and occupy similar microbial niches. *D. discoideum*'s ncDNA/tgDNA value falls in between those of the yeasts and those of *Plasmodium* species. This is interesting as *D. discoideum* has an involved life cycle, and is considered to have an evolutionary position close to the base of metazoan evolution [29], whereas malarial parasites are strictly unicellular and of seemingly lower complexity. However, the latter must survive in multiple hosts and circumvent the immune system of these hosts, and may therefore have higher functional complexity than their strict unicellularity would suggest. Likewise, another human parasite and the causative agent of sleeping sickness, *Tyrpanosoma brucei*, occupies the highest value for the unicellular eukaryotes. Trypanosomes are described as possessing a number of seemingly complex traits despite their unicellularity, including the ability to infect multiple hosts, a non-obligatory sexual cycle, hybrid genotypes, polycistronic transcription, and a developed drug resistance [30].

Multicellular eukaryotes also display consistent trends in their noncoding DNA composition, ranging from 0.62 for the fungus *Neurospora crassa*, 0.71 – 0.80 for plants, 0.74 – 0.93 for invertebrates, 0.87 for a urochordate, and 0.89 – 0.98 for vertebrates, which is broadly consistent with their relative developmental and functional complexity. The ncDNA/tgDNA ratio again helps to resolve previous incongruities in the gene number-complexity relationship in complex organisms. Both the fruit fly and urochordate have fewer protein-coding genes than the apparently less complex roundworm. In the case of the fruit fly, Adams *et al* [31] attempted to explain this by positing that even though *Drosophila* had fewer genes, these genes had comparable functional diversity to those in *C. elegan*s. The data presented here, however, shows that the ncDNA/tgDNA ratio increases from roundworm to fruit fly to urochordate consistent with the apparent increase in complexity. The ncDNA/tgDNA ratio of Neurospora places it at the base of the multicellular eukaryotes, consistent with previous assessments [32]. The vertebrates species examined have the highest average value and highest overall value. Additionally, the vertebrate sub-group shows an increase in ncDNA/tgDNA value from Fugu, to mouse, to human. This is noteworthy in view of the fact that all three of these species have nearly identical gene numbers.

While the ncDNA/tgDNA ratio offers the most insight of any gross genomic characteristic, it does introduce other apparent incongruities. For example Anopheles has a higher ncDNA/tgDNA ratio than the pufferfish, which may be explained by variable loads of repetitive sequences (see below). Therefore, these ratios should only be viewed as bandwidths generally indicative of broad trends. These trends are, of course, limited by the fact that few eukaryote genomes have been completely sequenced. We suspect that future sequencing projects and subsequent analysis of already sequenced genomes will facilitate a tightening of the median of these bandwidths.

*The C-value paradox*

Consideration of the significance of non-protein-coding DNA sequences in genomes has long been complicated and confused by the so-called C-value paradox [6, 9], which simply holds that the amount of genomic DNA does not correlate with organismal complexity, or at least that there appear to be significant exceptions to such correlations. For example, amphibians can have as much as 40 times as much DNA per cell than mammals [9], but are ostensibly no more complex. Such inconsistencies can be accounted for by variable amounts of repetitive DNA sequences, which can introduce some scatter in the calculated ncDNA/tgDNA ratio, although it does not significantly disturb the overall trends. We have not generally discounted or attempted any correction for such repetitive sequences in our overall analysis because of the different types and uncertain significance of these sequences, some of which may have acquired function or play an as yet undefined role in the organisms concerned. One can, however, make some general observations by analyzing data from genomes of organisms that have been sequenced, particularly those multicellular organisms that are relatively unencumbered by repetitive sequences, such as pufferfish.

In contrast to amphibians, the pufferfish has a small genome with only about 10% "repetitive" sequences. In this organism there are about 30,000 protein coding genes (similar to mammals), whose protein-coding sequences occupy approximately 11% of the total genome (calculated from the number of protein-coding genes and the average length of protein-coding sequences) [4]. The majority of the introns in these genes are small (and presumably vestigial), but many are large and collectively still exceed the amount of exonic sequence by a considerable margin. Since protein-coding genes account for about one third of the genome [4] the overall intron-exon ratio in *Fugu* is about 2:1, which means that introns of protein coding genes occupy about one fifth of the *Fugu* genome. This still leaves about two thirds of the genome containing highly unique sequences unaccounted for, which seems remarkable especially since there appears to be "rapid deletion of nonfunctional sequences" in this organism [4]. In fact, data supporting the possible role of unique noncoding sequence in Fugu has been revealed by a recent comparison of the Fugu Sonic Hedgehog (Shh) gene to mouse, human, and zebrafish Shh genes, which has shown the conservation of 19 noncoding elements [33]. Furthermore, comparison of the Fugu genome with those of human and mouse, which have about 40% repetitive sequences, shows that while the number of protein-coding genes is similar, the amount of noncoding sequences has risen, such that even after removing this 40% from consideration (for the sake of this argument), protein coding sequences only account for around 3% of the unique sequences in these genomes, a decrease in the ncDNA/tgDNA ratio to around 0.97, which is still very high. In fact, given that the numbers of protein coding genes (and the proportion of repetitive sequences) are similar in human and mouse, but the latter genome is somewhat smaller (2,500 Mb vs 2,900 Mb), the ncDNA/tgDNA ratio in mouse is less than that of human, consistent with the expanded skeleto-muscular and neural complexity in the latter.

*The role of noncoding DNA sequences in the programming of eukaryotic complexity*

In light of these observations, we propose that the amount of noncoding DNA in relation to genome size may be a more accurate indicator of relative biological complexity than the absolute number of protein-coding genes, although the latter will obviously contribute, especially in view of the expanded numbers of isoforms that may be produced by alternative splicing. Although both have some incongruities, the former is more consistent, and unambiguously delineates enucleate, nucleate, and multicellular nucleate species. The correlation between the amount of ncDNA and complexity clearly implies that these noncoding sequences play a major role in the genomic programming that may result in differential levels of biological complexity.

The amount of cis-acting regulatory sequences in the genomes of the higher organisms is impossible to assess at this time. However, it appears that the amount of trans-acting regulatory information that is produced has been very seriously underestimated, at least partly because of the assumption that this information is conveyed primarily by proteins. It is becoming clear that this is not the case. The vast majority (approximately 98%) of the transcriptional output of the human genome is noncoding RNA, which is by definition reflected in the ncDNA/tgDNA ratio. Additionally, at least half the human genome is actually transcribed, even though protein coding sequences only account for about 1.5% of these RNA sequences [34, 35]. Recent evidence has indicated that there are many thousands of ncRNA transcripts in mammals [16]. All those that have been studied exhibit tissue-specific expression and have been shown to play roles in germ cell formation, dosage compensation, neural, kidney and liver development, stress responses, immune cell activation, and in several diseases [7]. These may be just the tip of an iceberg as many complex genetic phenomena in the higher organisms are now known to be directed by, or connected to, RNA signaling [34, 36]. This suggests that any tally of protein coding "genes" is at best an incomplete measure of the sophistication of the genomic programming and complexity of an organism, and this measure should now be viewed as just one component of a much larger suite of genomic and genetic information.

The literature is replete with evidence suggesting that noncoding sequences are functional. It has been suggested that transposons and other repeat sequences may play varied roles in the evolution of a genome, including exon shuffling, homologous recombination and the divergence of transcriptional regulatory elements [37]. These roles may permit comparatively fast evolutionary changes in genome structure and the regulation and diversity of encoded gene products [37, 38]. It has also been suggested that repetitive elements may be components of key regulatory systems, and are evolving *in situ* as part of the overall genetic programming of the organism. This is supported by the fact that that many repetitive elements are expressed (as RNA), which suggests that they may be potential contributors to the trans-acting aspect of the regulatory architecture. Furthermore, repetitive elements have been implicated in a number of essential genetic activities including the formation of higher order nuclear structures, centromere formation, chromatin condensation, functioning as nucleation centers for methylation, cell proliferation, cell stress response, translation, binding cohesions to chromosomes, and DNA repair [39]. Introns themselves permit a more sophisticated genetic output through their ability to create alternate gene products via alternate splicing, regulation of gene expression, and the ability to generate new proteins by exon shuffling [40], as well by their potential ability to transmit RNA signals into the system [34, 35]. Intronic sequence conservation between humans, mice, dogs, whales, and seals shows a degree of conservation above that expected by chance and has revealed evolutionary constraints on noncoding sequence, which may have an impact on studies which have analyzed these sequences under the assumption of neutral selection [8, 41]. There are also non-intronic noncoding elements which are conserved across species which, as yet, have no known function. These include 2,262 of 3,491 conserved noncoding sequences between human chromosome 21 and a syntenic region in mouse, and noncoding sequences conserved between human chromosome 20 and mouse chromosome 2 with a degree of conservation as intense as coding sequence [42, 43]. It is also worth acknowledging the more general work that has been done suggesting that there are one or more structured languages residing in ncDNA using power law analysis, Zipf language analysis, Shannon analysis, and measures of redundancy [44-47]. Lastly, it is interesting to note that increased complexity is primarily associated with an expansion in the control system, rather than the functional components [48]. The control system of complex eukaryotes, particularly that which controls the precise trajectories of differentiation and development, may be far greater and encoded in ways other than regulatory proteins. In general complexity is a function of the amount of information required to specify the system, which in biological systems is believed to be encoded in the genome. Our results suggest that the majority of this information is likely to reside outside of protein-coding sequences, and that much more of the genomes of complex organisms are genetically active than previously thought.

## Conclusions

The data here show that the amount of noncoding DNA per genome is a more valid measure of the complexity of an organism than the number of protein-coding genes, and may be related to the emergence of a more sophisticated genomic or regulatory architecture [49, 50], rather than simply a more sophisticated proteome. Although the genomes of complex organisms are complex entities with many passengers, the ncDNA/tgDNA trend, in synthesis with data associated with both the known and theoretical importance and function of ncDNA, suggests that there is in fact much less, if any, “junk” in the genomes of the higher organisms than has previously been supposed. To the contrary, we suggest that a number of the elements in the noncoding portion of the genome may be central to the evolution and development of multicellular organisms. Understanding the functions of these non-protein-coding sequences, not just the functions of the proteins themselves, will be vital to understanding the genetics, biology and evolution of complex organisms.

## Materials and Methods

Information regarding the composition of each genome was acquired from the appropriate genome sequencing publication and any related literature. Noncoding DNA was considered DNA known or predicted to fit into one of the categories outlined in the *Analysis of sequenced genomes* section in Results and Discussion. Prokaryotic genomes were analyzed considering only the main chromosome. Plasmids were excluded from this analysis. A total of 76 prokaryotic genomes were randomly chosen to be examined out of a possible pool of 125. 67 were judged to have sufficient information for analysis. All eukaryote sequences with substantial genomic information were included in the analysis. Incomplete genomic information was allowed in two cases. The Dictyostelium ncDNA/tgDNA value is based on the sequence of chromosome 2. The trypansome ncDNA/tgDNA value is based on the sequence of chromosomes 1 and 2. References and details on publications used for each species can be found in Table 1. Additional data and frequently updated ncDNA/tgDNA values can be accessed at www.noncodingDNA.com.

**Acknowledgements**

We would like to thank Garret L. Yount for his support.

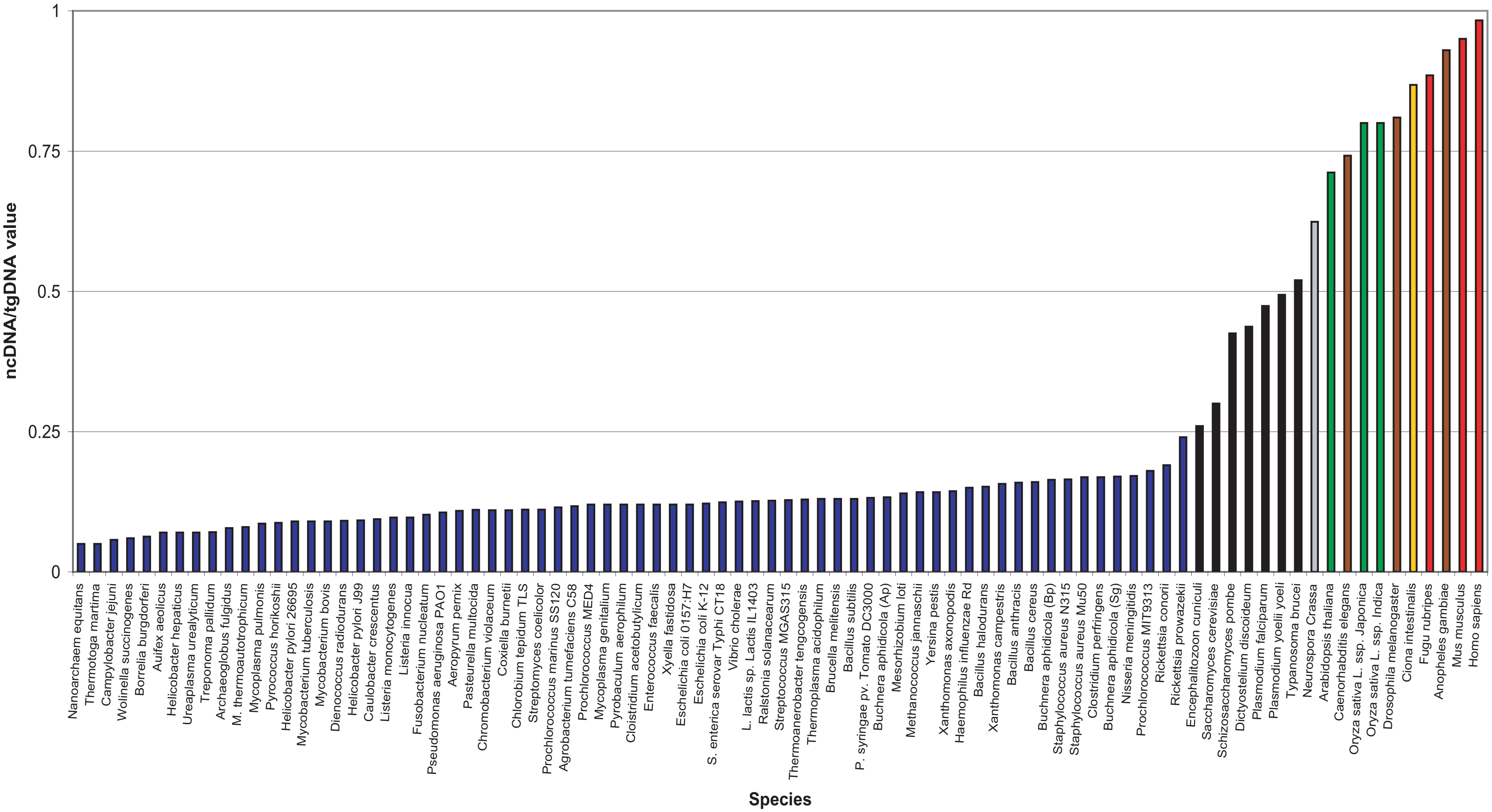

**Figure 1** - The increase in the ratio of noncoding DNA to total genomic DNA (ncDNA/tgDNA) is shown to correlate with increasing biological complexity. For ease of phylogenetic parsing, prokaryotes are labeled in blue, unicellular eukaryotes in black, the multicellular fungus *Neurospora crassa* in gray, plants in green, non-chordate invertebrates in brown, the urochordate *Ciona intestinalis* in orange, and vertebrates in red.

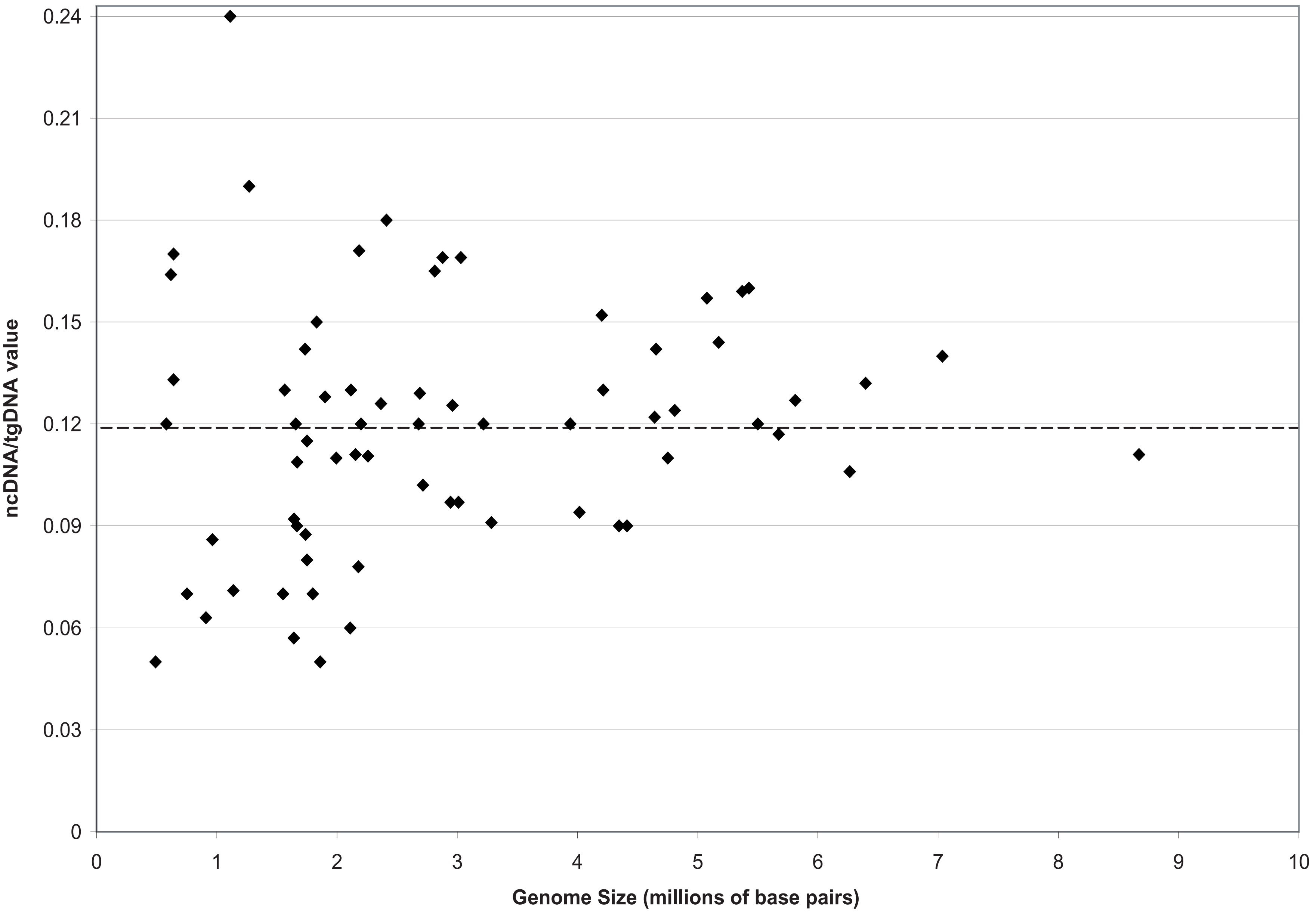

**Figure 2** - Prokaryotic species density by ncDNA/tgDNA value and genome size. The dashed line represents the ncDNA/tgDNA mean for prokaryotes, 0.1177. 67 bacterial species show a nonlinear ncDNA/tgDNA trend in relation to genome size ($r = 0.15$), and an unexpected density approaching the mean ($p = 1.58 \times 10^{-13}$).

**Table 1**: A survey of ncDNA/tgDNA values, gene number, and genome size

| Organism | ncDNA/tgDNA value | Gene number | Genome size | NcDNA/tgDNA calculation comments | References |
|---|---|---|---|---|---|
| *Nanoarchaeum equitans* | 0.05 | 552 | 490885 | | [51] |
| *Thermotoga martima* | 0.05 | 1877 | 1860725 | | [52] |
| *Campylobacter jejuni* | 0.057 | 1654 | 1641181 | | [53] |
| *Wolinella succinogenes* | 0.06 | 2046 | 2110355 | | [54] |
| *Borrelia burgdorferi* | 0.063 | 1283 | 1443725 | | [55] |
| *Auifex aeolicus* | 0.07 | 1512 | 1551335 | | [56] |
| *Helicobacter helaticus* | 0.07 | 1875 | 1799146 | | [57] |
| *Ureaplasma urealyticum* | 0.07 | ? | 751719 | | [58] |
| *Treponma pallidum* | 0.071 | 1041 | 1138006 | | [59] |
| *Archaeoglobus fulgidus* | 0.078 | 2436 | 2178400 | | [60] |
| *Methanobacterium thermoautotrophicum* | 0.08 | 1855 | 1751377 | | [61] |
| *Mycoplasma pulmonis* | 0.086 | 782 | 963879 | | [62] |
| *Pyrococcus horikoshii* | 0.0875 | 2061 | 1738505 | | [63] |
| *Mycobacterium tuberculosis* | 0.09 | 3924 | 4411529 | | [64] |
| *Mycobacterium bovis* | 0.09 | 3951 | 4345492 | | [65] |
| *Helicobacter pylori 26695* | 0.09 | 1590 | 1667867 | | [66] |
| *Dienococcus radiodurans* | 0.091 | 3187 | 3284156 | | [67] |
| *Helicobacter pylori J99* | 0.092 | 1495 | 1643831 | | [66] |
| *Caulobacter crescentus* | 0.094 | 3767 | 4016942 | | [68] |
| *Listeria monocytogenes* | 0.097 | 2853 | 2944528 | | [69] |
| *Listeria innocua* | 0.097 | 2973 | 3011209 | | [69] |
| *Fusobacterium nucleatum* | 0.102 | 2,067 | 2714500 | | [70] |
| *Pseudomonas aeruginosa* | .106 | 5570 | 6264403 | | [71] |
| *Aeropyrum pernix* | 0.1088 | 2618 | 1669695 | | [72] |
| *Coxiella burnetii* | 0.11 | 2094 | 1995275 | | [73] |
| *Chromobacterium violaceum* | 0.11 | 4431 | 4751080 | | [74] |
| *Pasteurella multocida* | 0.1106 | 2014 | 2257487 | | [75] |
| *Streptomyces coelicolor* | 0.111 | 7825 | 8670000 | | [76] |
| *Chlorobium tepidum TLS* | 0.111 | 2288 | 2154946 | | [77] |
| *Prochlorococcus marinus* | 0.115 | 1884 | 1751080 | | [78] |
| *Agrobacterium tumefaciens C58* | 0.117 | 5419 | 5674062 | | [79] |
| *Mycoplasma genitalium* | 0.12 | 470 | 580070 | | [80] |
| *Pyrobaculum aerophilum* | 0.12 | 2587 | 2200000 | | [81] |
| *Procholorococcus MED4* | 0.12 | 1716 | 1657990 | | [82] |
| *Cloistridium acetobutylicum* | 0.12 | 3740 | 3940880 | | [83] |
| *Enterococcus faecalis* | 0.12 | 3182 | 3218031 | | [84] |
| *Xylella fastidosa* | 0.12 | 2782 | 2679305 | | [85] |
| *Eschelichia coli 0157:H7* | 0.12 | 5447 | 5500000 | | [86] |
| *Eschelichia coli K-12* | 0.122 | 4288 | 4641000 | | [86] |
| *Salmonella enterica serovar Typhi CT18* | 0.124 | 4599 | 4809037 | | [87] |
| *Vibrio cholerae* | 0.1255 | 3885 | 4034065 | | [88] |
| *Lactococcus lactis sp. Lactis IL* | 0.126 | 2365589 | 2310 | | [89] |
| *1403* | | | | | |
| *Ralstonia solanacearum* | 0.127 | 5129 | 5810922 | | [90] |
| *Streptococcus MGAS315* | 0.128 | 1865 | 1900521 | | [91] |
| *Thermoanerobacter tengcogensis* | 0.129 | 2588 | 2689445 | | [92] |
| *Thermoplasma acidophilum* | 0.13 | 1509 | 1 564 905 | | [93] |
| *Brucella melitensis* | 0.13 | 3197 | 3 294 935 | | [94] |
| *Bacillus subtilis* | 0.13 | 4100 | 4 214 810 | | [95] |
| *Pseudomonas syringe pv. Tomato DC300* | 0.132 | 5615 | 6 397 126 | | [96] |
| *Buchnera aphidicola (Ap)* | 0.133 | 618 | 640000 | | [97] |
| *Methanococcus jannaschii* | 0.14 | 1738 | 1734000 | | [98] |
| *Mesorhizobium loti*[b] | 0.142 | 6752 | 7036071 | | [99] |
| *Yersina pestis* | 0.142 | 3908 | 4653728 | | [100] |
| *Xanthomonas axonopodis* | 0.144 | 2710 | 5175554 | | [101] |

| | | | | |
|---|---|---|---|---|
| *Haemophilus influenzae Rd* | 0.15 | 4553 | 4524893 | [102] |
| *Bacillus halodurans* | 0.152 | 4066 | 4202353 | [95] |
| *Xanthomonas campestris* | 0.157 | 2708 | 5076187 | [101] |
| *Bacillus anthracis* | 0.159 | 5842 | 5370060 | [103] |
| *Bacillus cereus* | 0.16 | 5366 | 546909 | [104] |
| *Buchnera aphidicola (Bp)* | 0.164 | 553 | 618000 | [105] |
| *Staphylococcus aureus N315* | 0.165 | 2595 | 2813641 | [106] |
| *Staphylococcus aureus Mu50* | 0.169 | 2687 | 2878084 | [106] |
| *Clostridium perfringens* | 0.169 | 2660 | 3031430 | [107] |
| *Buchnera aphidicola (Sg)* | 0.17 | 545 | 640000 | [97] |
| *Nisseria meningitidis* | 0.171 | 2121 | 2184406 | [53] |
| *Prochlorococcus MIT9313* | 0.18 | 2275 | 2410873 | [82] |
| *Rickettsia conorii* | 0.19 | 1374 | 1268755 | [108] |
| *Rickettsia prowazekii* | 0.24 | 834 | 1111523 | [108] |
| *Encephalitozoon cuniculi* | 0.26 | 1997 | 2507519 | [109] |
| *Saccharomyces cerevisiae* | 0.295 | 6000 | 12100000 | [110] |

| | | | | | |
|---|---|---|---|---|---|
| *Schizosaccharomyces pombe* | 0.425 | 4824 | 13800000 | | [111] |
| *Dictyostelium discoideum* | 0.437 | 11000 | 32000000 | Calculations based on analysis of chromosome 2. | [112] |
| *Plasmodium falciparum* | 0.474 | 5268 | 22853764 | | [113] |
| *Plasmodium yoelii yoelii* | 0.494 | 5878 | 23100000 | | [114] |
| *Trypanosoma brucei*** | 0.52 | 687 | 2258403 | Calculations based on chromomsomes I and II (~10% of the genome). | [115] [116] |
| *Neurospora crassa* | 0.624 | 10000 | 40000000 | | [117] |
| *Arabidopsis thaliana* | 0.712 | 25000 | 115409949 | | [118] |
| *Caenorhabditis elegans* | 0.7419 | 19049 | 97000000 | | [119] |
| *Orya sativa* L. ssp. *japonica* | 0.80 | 32,000 – 50,000 | 420000000 | Calculations based on genome draft estimates and data gleaned from chromosome 1 and 4 sequences. | [120-122] |
| *Oryza sativa* L. ssp. *indica* | 0.80 | 46022 – 55615 | 466000000 | See notes for *O. japonica* | [121-123] |
| *Drosophila melanogaster* | 0.81 | 13600 | 120000000 | | [124, 125] |
| *Ciona Intestinalis* | 0.868 | 16000 | 160000000 | | [126, 127] |
| *Fugu rubripes* | 0.885 | 31,059 | 365000000 | Calculations include known intergenic sections and intronic DNA estimates. | [128] |
| *Anopheles gambiae* | 0.93 | 13683 | 278244063 | | [125] |
| *Mus musculus* | 0.95 | 37000 | 2500000000 | Calculations of ratio estimate based on mouse draft genome sequence. | [129] |
| *Homo sapiens* | 0.983 | 30000 | 3000000000 | Calculations based on estimated number of genes and gene size. | [130, 131] |

**Table References**